About the Structure of a General Equation of State for Liquids and Gases

I.H. Umirzakov

## Abstract

It is shown that the functional form of the equation of state of [4-8] is not correct in general case.

The equation of state (the dependence of a pressure $p(T,v)$ or compressibility factor $z(T,v)$ on temperature $T$ and molar volume $v$) can be obtained from the partition function known [1]. However the partition function of bulk system is known only for the ideal gas and for the system of harmonic oscillators [2-3]. Therefore in order to establish the equation of state (EOS) it is necessary to use experimental $(p,v,T)$-data and other thermodynamic data for real substances [4], and the functional form of the EOS based on equilibrium statistical mechanics. The functional form of the EOS of [5-8] is not correct in general case.

The equation

$$\frac{\partial z(T,v)}{\partial T} = \frac{v}{RT^2} \cdot \frac{\partial U(T,v)}{\partial v}, \qquad (1)$$

where $R$ is the universal gas constant, $z(T,v) \equiv \frac{p(T,v) \cdot v}{RT}$ is the compressibility factor, and $U(T,v)$ is the internal energy, can be obtained [5-8] from thermodynamic equation [1]

$$\frac{\partial U(T,v)}{\partial v} = T \cdot \frac{\partial p(T,v)}{\partial T} - p(T,v).$$

We integrate the equation (1) and obtain

$$z(T,v) = \int_{T_0(v)}^{T} \frac{v}{RT'^2} \cdot \frac{\partial U(T',v)}{\partial v} dT' + z(T_0(v),v), \qquad (2)$$

where $T_0(v)$ and $z(T_0(v),v)$ are the arbitrary functions.

We have [9]

$$\int_{T_0(v)}^{T} \frac{v}{RT'^2} \cdot \frac{\partial U(T',v)}{\partial v} dT' = F(v,T) - F(v,T_0(v)),$$

where

$$\frac{\partial F(v,T)}{\partial v} = \frac{v}{RT^2} \cdot \frac{\partial U(T,v)}{\partial v}.$$

The Eq. (2) can be presented as

$$z(T,v) = h(T,v) + z_0(v), \tag{3}$$

where $h(T,v) = F(v,T) - F(v,T_0(v))$, $z_0(v) \equiv z(T_0(v),v)$.

It is implicitly assumed in [5-8] that in (3)

$$F(v,T_0(v)) \equiv 0, \tag{4}$$

$$z_0(v) \neq 0. \tag{5}$$

However there is no proofs in [5-8] that the $F(v,T_0(v)) \equiv 0$ for arbitrary functions $T_0(v)$ and $U(T,v)$, and the identity $z_0(v) \equiv 0$ is not correct.

If

$$\left| \int_\infty^T \frac{v}{RT'^2} \cdot \frac{\partial U(T',v)}{\partial v} dT' \right| < \infty, \tag{6}$$

then the equation (2) can be presented as

$$z(T,v) = \int_\infty^T \frac{v}{RT'^2} \cdot \frac{\partial U(T',v)}{\partial v} dT' + z_0(v) = F(v,T) - F(v,\infty) + z_0(v), \tag{7}$$

where $z_0(v) = z(T,v)|_{T \to \infty}$.

However there is no proofs in [5-8] that the $F(v,\infty) \equiv 0$ for arbitrary function $U(T,v)$, and the identity $z_0(v) \equiv 0$ is not correct in (7).

One can easily see that the inequality (6) is not valid for the EOS

$$p(T,v) = kT \cdot f(v) \cdot q(T,v), \tag{8}$$

where $q(T,v)$ is an arbitrary function, $q(T,v) \neq \chi(v)$ for $0 < T < \infty$, where $\chi(v)$ is an arbitrary function, and $q(T,v)|_{v \to \infty} = 1$. Particularly, $f(v) = 1/(v-b)$, $q(T,v) = \exp(-a/vkT)$, where $a = const > 0$, $b = const > 0$, for the Dieterici equation of state [10,11], and $f(v) = [v^3 + v^2 b_0 + v b_0^2 - b_0^3]/v(v-b_0)^3$, $b_0 = b/4$, $q(T,v) = \exp(-a/vkT)$ for the Carnahan-Starling-Dieterici equation of state [12].

There is no proof in [5-8] that the EOS cannot have the form (8).

The virial equation of state [1,11,13] gives

$$z(T,v) = 1 + \sum_{n=2}^\infty B_n(T)/v^{n-1}, \tag{9}$$

where $B_n(T)$ is the $n$-th virial coefficient. One can put in (3) that $h(T,v) = 1 + \sum_{n=2}^{\infty} B_n(T)/v^{n-1}$ and $z_0(v) = 0$. Therefore the equation of state in the form (3)-(5) is not valid.

There is no proof in [5-8] that the EOS cannot have the form of the virial EOS (9).

One can show that the inequality (6) is not valid for the special form of the EOS

$$U(T,v) = T^{\lambda(v)} \cdot f(v) + g(T,v), \qquad (10)$$

where $\lambda(v)$, $f(v)$ and $g(T,v)$ are arbitrary functions, if $\lambda(v) \geq 1$.

There is no proof in [5-8] that the EOS cannot have the special form (10). There is no proof in [5-8] that the inequality (6) takes place in general case.

We see from above consideration that the form (3)-(5) of the equation of state is not correct for the general case. The validity of the form of the equation of state (3)-(5) cannot be proved in the framework of the equilibrium thermodynamics. The functional form (structure) of the EOS can be established in the framework of the equilibrium statistical mechanics.